\documentclass[reqno]{amsart}
\usepackage{hyperref}
\usepackage{amsmath,amssymb} 
\usepackage{graphicx}
\DeclareGraphicsExtensions{.pdf,.png,.jpg}

\usepackage[T2A]{fontenc}
\usepackage[utf8]{inputenc}
\usepackage{textcomp}

\allowdisplaybreaks

\theoremstyle{plain}
\newtheorem{theorem}{Theorem}[section]

\theoremstyle{definition}

\theoremstyle{remark}

\numberwithin{equation}{section}

\begin{document}
		\title[ The exact solution of the Axial Next - Nearest - Neighbor Ising model on width 2 spin chains
	]
	{The exact solution of the Axial Next - Nearest - Neighbor Ising model on width 2 spin chains
	}

	\author[P. Khrapov]{Pavel Khrapov}
	\address{Pavel Khrapov \\ Department of Mathematics
		\\ Bauman Moscow State Technical University \\  ul. Baumanskaya 2-ya, 5/1, Moscow \\ 105005, Moscow,  Russian Federation}  
	\email{khrapov@bmstu.ru , pvkhrapov@gmail.com }
	
	\author[A. Loginova]{Anastasia Loginova}
	\address{Anastasia Loginova \\ Department of Mathematics
		\\ Bauman Moscow State Technical University \\  ul. Baumanskaya 2-ya, 5/1, Moscow \\ 105005, Moscow,  Russian Federation}  
	\email{loginovaaa@student.bmstu.ru, nastya.loginova@mail.ru}	
	
	
	
	\begin{abstract}
		Using the transfer matrix method for Axial Next-Nearest Neighbor Ising model without an external field on a closed chain of spins of width 2 in the direction of interaction of only nearest neighbors and length $L$ in the direction of interaction of nearest neighbors and next nearest neighbors, we found
		exact values of the partition function in a finite strip of length $L$ , free energy, internal energy per node, heat capacity in a finite strip of length $L$ and in the thermodynamic limit at $L \to \infty $. The entire spectrum of the $16 \times $16 transfer matrix and the structure of all transfer matrix eigenvectors are found. The problem of finding the spectrum of the transfer matrix is reduced to finding the spectrum of two matrixes of size $ 8 \times 8 $. The characteristic polynomial of the first matrix is decomposed into two linear polynomials of multiplicity two and two square trinomials. The characteristic polynomial of the second matrix is decomposed into two polynomials of the fourth degree, the roots of which are found by the Ferrari method. The free energy in the thermodynamic limit has a very simple form and is expressed in terms of the logarithm of the root of the quadratic equation.
		Two theorems on the physical characteristics of the ANNNI model in a finite closed strip of length $L$ and in the thermodynamic limit as $L \to \infty $ are formulated and proved. An example of calculation is given for some values of the model parameters in the thermodynamic limit, heat capacity graphs are plotted. A comparative analysis of the heat capacity for the double-stranded ANNNI model and the ANNNI model on the entire flat lattice is given. The closest connection and similarity of the physical characteristics of the two-chain model and the model on the entire flat lattice is shown.
	\end{abstract}
	
	\maketitle
	\tableofcontents

	\section{Introduction} 
		
	The abbreviation ANNNI stands for "Axial Next-Nearest Neighbor Ising model". In the ANNNI model, interactions bind spins at nearest and next nearest neighbors along one of the crystallographic axes of the lattice. The model is the prototype of complex spatially modulated magnetic superstructures in crystals.
		
	This model was introduced by Roger Elliott, a British theoretical physicist at the University of Oxford, in 1961 \cite{Elliott} to explain the existence of modulated phases in rare earth compounds. However, it was not until 1980 that the British scientist Michael Ellis Fischer and the German scientist Walter Selke gave the ANNNI model its name \cite{Fisher_Selke}. In their work, the authors considered the simplest anisotropic Ising model with the next nearest neighbor along one axis in the low-temperature region.
				
	Subsequently, the ANNNI model has been well researched. One of the most important works on this topic is \cite{Selke_1988}, which analyzes the ANNNI model in one, two and three dimensions.  The model has been studied both in the low-temperature and high-temperature regions: M.E. Fisher and W. Selke analyzed the model in the low-temperature region near its multiphase point \cite{Fisher_Selke_1981}, J. Oitmaa investigated the model in the high-temperature region in two- and three-dimensional spaces \cite{Oitmaa_1999}. The article \cite{Selke_Duxbury_1984} is devoted to the study of the mean field equations on a simple cubic
	ring model on finite lattices.  
	The authors of \cite{Sinai}, Ya. G. Sinai and E. I. Dinaburg, explored  the three-dimensional ANNNI model and derived a converging expression for the curve of the coexistence of different phases for low temperatures using a new extension of the Peierls contour method.

	The phase diagram of the two-dimensional ANNNI model is studied with CVM in \cite{Finel_de Fontaine_1986}. In \cite{Fisher_Szpilka_1987_1} a general formalism for the analysis of physical systems is developed, which is further applied to the ANNNI model in \cite{Fisher_Szpilka_1987_2}. The article \cite{Surda_2004} describes the developed effective-field method for calculating thermodynamic properties for the ANNNI model on a simple three-dimensional cubic lattice. In \cite{Gendiar_Nishino_2005}, the 3D ANNNI model is studied using a modified tensor product variational approach (TPVA).
		
	Most of the model studies were performed using the numerical Monte Carlo method. In \cite{Selke_MC_1981} a two-dimensional finite-size ANNNI model with periodic boundary conditions was studied by the Monte Carlo method. In \cite{Kaski_Selke_1985} the 3D ANNNI model is considered using the Monte Carlo method.The papers \cite{Zhang_Charbonneau_2010},\cite{Zhang_Charbonneau_2011} describe the developed modeling method based on thermodynamic integration, which avoids the expansion of the metastability of modulated phases and allows one to obtain the phase diagram of the canonical three-dimensional axial nearest neighbor Ising model. The authors of the article \cite{Da Silva_Alves_Drugowich de Felicio_2013} studied the critical behavior of second order points, in particular the Lifshitz point of the 3D ANNNI model, using the Monte Carlo method. The series of papers \cite {Murtazaev_Ibaev_2010}, \cite {Murtazaev_Ibaev_2011}, \cite {Murtazaev_Ibaev_2012_1}, \cite {Murtazaev_Ibaev_2012_2}, \cite {Murtazaev_Ibaev_2016}, \cite {Murtazaev_Ibaev_2018_1}, \cite {Murtazaev_Ibaev_Abuev_2018} is dedicated to the study of two-dimensional model ANNNI by Monte -Carlo based on modifications of the Metropolis algorithm, as well as its thermodynamic parameters.
	
	Another part of the work explores the ANNNI model using the construction of transfer matrices. In article \cite{Bhattacharyya_Dasgupta_1991} analysis of the exact solution for the statistical mechanics of the one-dimensional ferromagnetic ANNNI chain under an external magnetic field by the transfer matrix method is considered. The study of the two-dimensional ANNNI model using the transfer matrix was performed in the works \cite{Pesch_Kroemer_1985}, \cite{Hu_Charbonneau_2021}. The paper \cite{Pesch_Kroemer_1985} considers a toroidal model with $M$ rows and $N$ columns, where the number of rows tends to infinity. In \cite{Kassan-Ogly_Murtazaev_2014} phase transitions and magnetic properties of the Ising model with double and triple neighboring interactions are studied by the Monte Carlo method and the transfer matrix method.
		
	In this paper, using the transfer matrix method, similar to the transfer matrix method in \cite{Khrapov1}, \cite{Khrapov2}, for the ANNNI model without an external field for a closed chain of spins of width 2 and length $L$, the exact values of the partition function are found in a finite band of length $L$ , the exact values of free energy per site, internal energy per site, heat capacity are found in a finite band of length $L$ and in the thermodynamic limit. The free energy in the thermodynamic limit has a very simple form and is expressed in terms of the logarithm of the root of the quadratic equation. The transfer matrix moves in the direction of interaction of nearest neighbors and next nearest neighbors in contrast to the works \cite{Hu_Charbonneau_2021}, where the transfer matrix is considered in the direction of interaction of only nearest neighbors, with a finite number of spins in the direction of interaction of nearest neighbors and next nearest neighbors.
	An example of calculation is given for some values of the model parameters in the thermodynamic limit, heat capacity graphs are plotted. A comparative analysis of the heat capacity for the ANNNI model model on width 2 spin chains and the ANNNI model on the entire flat lattice is given. The closest connection and similarity of the physical characteristics of the two-chain model and the model on the entire flat lattice is shown.

	\section{Model description and main results} 
	\label{Model description and main results}
	
	\begin{figure} [h]
	\centering
	\includegraphics [scale = 0.5] {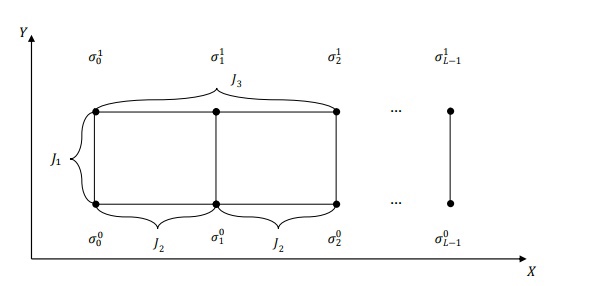}
	\caption{ANNNI model on width 2 spin chains}
	\end{figure}
	
	Let us consider a closed chain of spins of width 2 and length $L$ (Fig.1). The total number of the lattice sites is  $N=2 L$. We will assume that a particle is located at each site. The state of the particle is determined by the value of the spin $\sigma_{i}^{j}, i=0,1,...,L-1, j=0,1$ , which can take two values: $+1$ or $-1$ . Wherein $\sigma_{L}^{j} \equiv \sigma_{0}^{j}, \sigma_{L+1}^{j} \equiv \sigma_{1}^{j}, j=0,1. $ \\
	
	The Hamiltonian of the considered model has the form (Fig. 1):
	\begin{equation}
	\begin{gathered}
		\mathcal {H}_{L}
		= -\sum_{i=0}^{L-1} (J_{1}\sigma_{i}^{0}\sigma_{i}^{1} + J_{2}\sigma_{i}^{0}\sigma_{i+1}^{0} + J_{2}\sigma_{i}^{1}\sigma_{i+1}^{1} + J_{3}\sigma_{i}^{0}\sigma_{i+2}^{0} + J_{3}\sigma_{i}^{1}\sigma_{i+2}^{1}),   
	\end{gathered} 
	\end{equation}
	where $J_{m}$, $m=1,2,3$ are corresponding coefficients of interspin interaction. Thus,the model we consider is a closed ANNNI model on width 2 spin chainswithout an external magnetic field.
	
	The partition function can be written in the following form:	
	\begin{equation}
	\begin{gathered}
		Z_{{L}}=\sum_{\sigma}\exp ( -\mathcal {H}_{L}/{(k_B T)})=\\
		\sum_{\sigma}\exp ( \sum_{i=0}^{L-1} \frac {J_{1}}{k_{B}T}\sigma_{i}^{0}\sigma_{i}^{1} + \frac {J_{2}}{k_{B}T}\sigma_{i}^{0}\sigma_{i+1}^{0} + \frac {J_{2}}{k_{B}T}\sigma_{i}^{1}\sigma_{i+1}^{1} +\\ +\frac {J_{3}}{k_{B}T}\sigma_{i}^{0}\sigma_{i+2}^{0} + \frac {J_{3}}{k_{B}T}\sigma_{i}^{1}\sigma_{i+2}^{1}), 
	\end{gathered} 
	\end{equation}
	where  $k_{B}$ is Boltzmann's constant and summation perfomed over all spins. 
	
	To find the partition function, we introduce a transfer matrix, which is completely different from the transfer matrix considered in \cite{Hu_Charbonneau_2021}. The nonzero elements of the $16 \times 16$ transfer matrix $\Theta$ are as follows:
	\begin{equation}\label{Theta_no_zero_1}
	\begin{gathered}
		\Theta_{\{(\sigma_{i}^{0},\sigma_{i}^{1},\sigma_{i+1}^{0},\sigma_{i+1}^{1}), (\sigma_{i+1}^{0},\sigma_{i+1}^{1},\sigma_{i+2}^{0},\sigma_{i+2}^{1})\}} = \Theta_{k,l}=\\
		=\exp (\frac {J_{1}}{k_{B}T}\sigma_{i}^{0}\sigma_{i}^{1} + \frac {J_{2}}{k_{B}T}\sigma_{i}^{0}\sigma_{i+1}^{0} + \frac {J_{2}}{k_{B}T}\sigma_{i}^{1}\sigma_{i+1}^{1}  +\\ +\frac {J_{3}}{k_{B}T}\sigma_{i}^{0}\sigma_{i+2}^{0} + \frac {J_{3}}{k_{B}T}\sigma_{i}^{1}\sigma_{i+2}^{1}), \\
		k = (1-\sigma_i^0)/2+(1-\sigma_i^1)+ 2 (1-\sigma_{i+1}^0) + 4 (1-\sigma_{i+1}^1), \\
		l = (1-\sigma_{i+1}^0)/2+(1-\sigma_{i+1}^1)+ 2 (1-\sigma_{i+2}^0) + 4 (1-\sigma_{i+2}^1) . 
		\end{gathered} 
	\end{equation}

  	Then
  	\begin{equation}
  		\begin{gathered}
  			Z_{{L}}=Tr(	\Theta^L) , 
  		\end{gathered} 
  	\end{equation}

	Let us rewrite the expression (\ref{Theta_no_zero_1}) in the following form \cite{Khrapov1}, \cite{Khrapov2}:
	\begin{equation}\label{Theta_no_zero2}
	\begin{gathered}
		\Theta_{\{(\sigma_{i}^{0},\sigma_{i}^{1},\sigma_{i+1}^{0},\sigma_{i+1}^{1}), (\sigma_{i+1}^{0},\sigma_{i+1}^{1},\sigma_{i+2}^{0},\sigma_{i+2}^{1})\}} = \\ 
		(p^{\sigma_{i+1}^{0}\sigma_{i+1}^{1}}) \cdot (q^{\sigma_{i}^{0}\sigma_{i+1}^{0}}) \cdot (q^{\sigma_{i}^{1}\sigma_{i+1}^{1}}) \cdot (r^{\sigma_{i}^{0}\sigma_{i+2}^{0}}) \cdot (r^{\sigma_{i}^{1}\sigma_{i+2}^{1}}),
	\end{gathered} 
	\end{equation}
	where $p= \exp (\frac {J_{1}}{k_{B}T})$ , $q= \exp (\frac {J_{2}}{k_{B}T})$ , $r= \exp (\frac {J_{3}}{k_{B}T})$.
	
	The free energy of the system per one lattice site is determined in the standard way (\cite{Baxter}): 
	\begin{equation}
		\begin{gathered}
			f(T)=-k_{B}T\lim\limits_{N \to \infty} N^{-1} (\ln Z_{L}(T)),
		\end{gathered} 
	\end{equation}
	where $N$ is the number of sites of the considered lattice. \\
	
	The internal energy per one lattice site is equal to (\cite{Baxter}):
	\begin{equation}
		\begin{gathered}
			u(T)=-T^2 \frac{d}{d \mathrm{T}} [f(T)/T].
		\end{gathered} 
	\end{equation}
	
	The heat capacity per one lattice site is defined as (\cite{Baxter}):
	\begin{equation}
		\begin{gathered}
			C(T)=\frac{d}{d \mathrm{T}}	u(T) = -[2T \frac{d}{d \mathrm{T}} (f(T)/T)+ T^2 \frac{d^ 2}{d \mathrm{T^2}} (f(T)/T)].
		\end{gathered} 
	\end{equation}

	\begin{theorem} \label{t1}	
	\textbf{Main theorem}
	
	In the thermodynamic limit, free energy, internal energy and heat capacity are calculated in the following form:
		\begin{equation}
			\begin{gathered} \label{f}
				f(T)=-k_{B}T((\ln \lambda_{max} (T)) /2),
			\end{gathered} 
		\end{equation}		
		
		\begin{equation}
			\begin{gathered} \label{u}
				\frac {u(T)} {k_{B}}=-T^2 \frac{d}{d \mathrm{T}} [(\ln \lambda_{max} (T)) /2],
			\end{gathered} 
		\end{equation}
		
		\begin{equation}
			\begin{gathered} \label{c}
				\frac {C(T)} {k_{B}} =2T \frac{d}{d \mathrm{T}} ((\ln \lambda_{max} (T)) /2)+ T^2 \frac{d ^ 2}{d \mathrm{T^2}} ((\ln \lambda_{max} (T)) /2),				
			\end{gathered} 
		\end{equation}	
	
	where the largest eigenvalue $\lambda_{max}$ of the transfer matrix $\Theta$ has the form:
	\begin{center}
		\begin{equation} 
			\begin{gathered} \label{lambda_max_theorem}
				\lambda_{max} = \frac{1}{4p q^{2} r^{4}} (r^6(1+p^2)(1+q^4) + r^2  [(-1+p^2)^2 r^8 + \\(-1 + p^2)^2 q^{8} r^8   
				+ q^{4} (4-2r^8-2p^4(-2+r^8) + 4p^2(2+r^8))]^{\frac {1} {2}} + \\
				+[r^{4} (-16 p^{2} q^{4} + 32 p^2 q^4 r^4 - 16 p^2 q^4 r^8  
				+ ((1 + p^2) (1+q^4) r^4 + \\
				+[(-1 + p^2)^2 r^8 + (-1 + p^2)^2 q^8 r^8 + q^4 (4-2r^8-2p^4(-2+r^8) + 4p^2(2+r^8))]^{\frac {1} {2}}))^2]^{\frac {1} {2}}). 
			\end{gathered}
		\end{equation}
	\end{center}	 			
	\end{theorem}

  	To formulate the following theorem, we introduce the quantities $h_{1}, h_{2}$:  
	\begin{equation}\label{Theta_no_zero1}
	\begin{gathered}
  	h_{1} = -\frac{1}{2} (r^6(1+p^2)(1+q^4) - r^2 [4 q^4 + 8 p^2 q^4 + 4 p^4 q^4 + r^8 - \\
	- 2 p^2 r^8 + p^4 r^8 - 2 q^4 r^8 + 4 p^2 q^4 r^8 - 2 p^4 q^4 r^8 + 
	q^{8} r^8 - 2 p^2 q^{8} r^8 + p^4 q^{8} r^8]^{\frac {1} {2}} ), \\ 
	h_{2} =	-\frac{1}{2} (r^6(1+p^2)(1+q^4) + r^2 [4 q^4 + 8 p^2 q^4 + 4 p^4 q^4 + r^8 - \\
	- 2 p^2 r^8 + p^4 r^8 - 2 q^4 r^8 + 4 p^2 q^4 r^8 - 2 p^4 q^4 r^8 + 
	q^{8} r^8 - 2 p^2 q^{8} r^8 + p^4 q^{8} r^8]^{\frac {1} {2}} ) . 
	\end{gathered} 
	\end{equation}
		  
	\begin{theorem} \label{t2}
		
		The partition function in a finite closed strip of length $L$ can be written as:
		\begin{equation}
			\begin{gathered}
				Z_{L}=\sum_{i=1}^{16} \lambda_{i}^{L},
			\end{gathered} 
		\end{equation}
	
		where  $\lambda_{i} , i=1,...,16$ are the roots of the characteristic polynomial of the transfer matrix   $\Theta$. Wherein
		
			\begin{center}
			\begin{equation}
				\begin{gathered}
					\lambda_{1} =  \frac{-h_{2}+\sqrt{h_{2}^{2}-4 p^{2} q^{4} r^{4}(r^{4}-1)^{2}}} {2 p q^{2} r^{4}},
				\end{gathered}
			\end{equation}
		\end{center}	
		
		\begin{center}
			\begin{equation}
				\begin{gathered}
					\lambda_{2} = \frac{-h_{2}-\sqrt{h_{2}^{2}-4 p^{2} q^{4} r^{4}(r^{4}-1)^{2}}} {2 p q^{2} r^{4}},
				\end{gathered}
			\end{equation}
		\end{center}	
		
		\begin{center}
			\begin{equation}
				\begin{gathered}
					\lambda_{3} = \frac{-h_{1}+\sqrt{h_{1}^{2}-4 p^{2} q^{4} r^{4}(r^{4}-1)^{2}}} {2 p q^{2} r^{4}},
				\end{gathered}
			\end{equation}
		\end{center}	
		
		\begin{center}
			\begin{equation}
				\begin{gathered}
					\lambda_{4} = \frac{-h_{1}-\sqrt{h_{1}^{2}-4 p^{2} q^{4} r^{4}(r^{4}-1)^{2}}} {2 p q^{2} r^{4}}, 
				\end{gathered}
			\end{equation}
		\end{center}	
		
		\begin{center}
			\begin{equation}
				\begin{gathered}
					\lambda_{5,6} =\frac{r^{4}-1}{r^{2}}, 
				\end{gathered}
			\end{equation}
		\end{center}	
		
		\begin{center}
			\begin{equation}
				\begin{gathered}
					\lambda_{7,8} =\frac{1-r^{4}}{r^{2}}, 
				\end{gathered}
			\end{equation}
		\end{center} 
	
	The remaining $\lambda_{j}, j=9,...,16$ are the roots of two 4th degree equations
		\begin{center}
		\begin{equation} \label{poly_4_Theta3_1}
			\begin{gathered}
				(a_{2} x^{4}+b_{2} x^{3} + c_{2} x^{2} + d_{2} x + e_{2}) =0,
			\end{gathered} 
		\end{equation}
		\end{center}	

	\begin{center}
		\begin{equation} \label{poly_4_Theta3_2}
			\begin{gathered}			
				(a_{3} x^{4}+b_{3} x^{3} + c_{3} x^{2} + d_{3} x + e_{3}) =0,
			\end{gathered} 
		\end{equation}
	\end{center}
	
		where
		$$a_{2} = q^2 r^8,  b_{2} = p r^{10} - p q^4 r^{10}, 
		c_{2} = q^2 r^4 + p^2 q^2 r^4 - q^2 r^{12} - p^2 q^2 r^{12}, $$
		$$d_{2} = -p r^6 + p q^4 r^6 + 2 p r^{10} - 2 p q^4 r^{10} - p r^{14} + p q^4 r^{14},$$
		$$ e_{2} = p^2 q^2 (-1 + r)^4 (1 + r)^4 (1 + r^2)^4,$$
		$$a_{3} = p^2 q^2 r^8,  b_{3} = p r^{10} - p q^4 r^{10}, 
		c_{3} = q^2 r^4 + p^2 q^2 r^4 - q^2 r^{12} - p^2 q^2 r^{12}, $$
		$$d_{3} = -p r^6 + p q^4 r^6 + 2 p r^{10} - 2 p q^4 r^{10} - p r^{14} + p q^4 r^{14},$$
		$$ e_{3} = q^2 (-1 + r)^4 (1 + r)^4 (1 + r^2)^4,$$
		The roots of the equations (\ref{poly_4_Theta3_1}) and (\ref{poly_4_Theta3_2}) are found by the Ferrari method described in the \ref{appendix}.
		
	\end{theorem}
	Theorems (\ref{t1}), (\ref{t2}) are proved in the section \ref{Proof of theorems}.

	\section{Example} 
	\label{Example} 
	Let us apply the formulas (\ref{f}) - (\ref{c}) to find the thermodynamic characteristics of the ANNNI model for $J_1/k_{B}=1, J_2/k_{B}=1/2, T \in [0.05,4], J_3/k_{B} \in [0,2] $ (fig. 2 and fig. 3). The heat capacity graphs are shown below. Let us note that the results obtained are similar to those in (\cite{Hu_Charbonneau_2021}):  the numerically phase diagram of the two-dimensional ANNNI model in (\cite{Hu_Charbonneau_2021}) calculated using the transfer matrix and the heat capacity calculated analytically using the transfer matrix show that the phase transition is observed at $J_3/k_{B}=0.5$ at the low temperature region.
	  
	\begin{figure} [h]
	\centering
	\includegraphics [scale = 0.3] {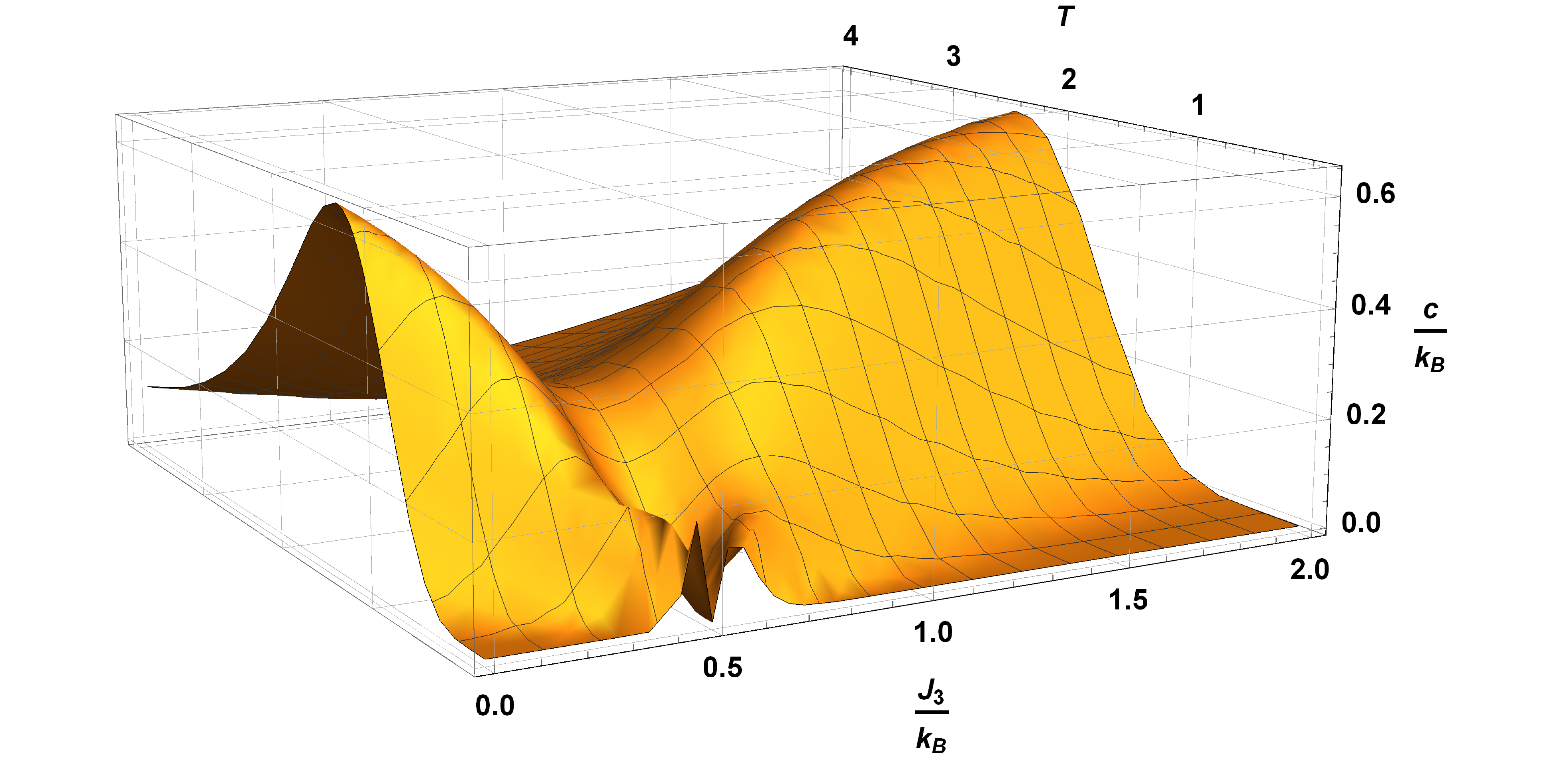}
	\caption{The heat capacity graph for $J_1/k_{B}=1, J_2/k_{B}=1/2, T \in [0,4], J_3/k_{B} \in [0,2]$}
	\end{figure}

	\begin{figure} [h!]
	\centering
	\includegraphics [scale = 0.5] {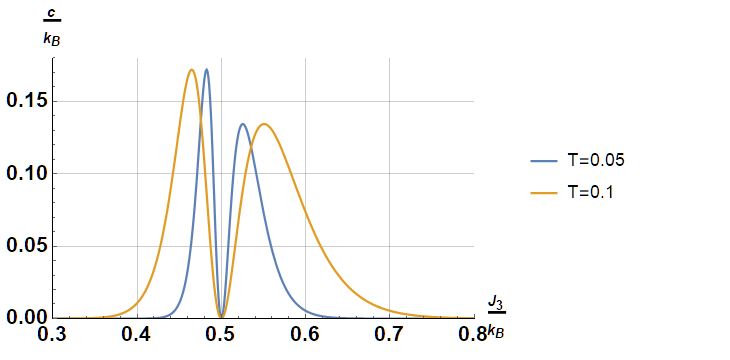}
	\caption{The heat capacity graph for $T=0.05, T=0.1, J_3/k_{B} \in [0,1]$}
	\end{figure}

	\section{Proof of theorems} 
	\label{Proof of theorems} 
	
	Let us find the eigenvalues of the transfer matrix $\Theta$ to prove the theorems \ref{t1}, \ref{t2} . 

 	The transfer matrix  $\Theta$, matrix that flips all spins $\Phi$,	
	$$
	\Phi_{\{(\sigma_{0}^{0},\sigma_{0}^{1},\sigma_{1}^{0},\sigma_{1}^{1}), (\zeta_{0}^{0},\zeta_{0}^{1},\zeta_{1}^{0},\zeta_{1}^{1})\}} =\left\{ 
	\begin{array}
	[c]{ll}
	1, if  \;\; \sigma_{i}^{j}=-\zeta_{i}^{j}, i=0,1, j=0,1
	\\
	0, in \; another \; cases	
	\end{array}
	\right  \} 
	$$ 	 	
 	and permutation matrix $\Gamma$,
 
  	$$
 	 \Gamma_{\{(\sigma_{0}^{0},\sigma_{0}^{1},\sigma_{1}^{0},\sigma_{1}^{1}), (\zeta_{0}^{0},\zeta_{0}^{1},\zeta_{1}^{0},\zeta_{1}^{1})\}} =\left\{
 	 	\begin{array}
 	 		[c]{ll}
 	 		1, if  \;\; \sigma_{i}^{j}=\zeta_{i}^{1-j}, i=0,1, j=0,1
 	 		\\
 	 		0, in \; another \; cases	
 	 	\end{array}
 	 	\right  \} 
	 	$$ 	 	
 	commute with each other.

 	The eigenvalues of the matrix $\Phi$ are
 	\begin{equation}
		\begin{gathered}
     		\{-1, -1, -1, -1, -1, -1, -1, -1, 1, 1, 1, 1, 1, 1, 1, 1\}			.		
		\end{gathered}  	 
	\end{equation} 	 
 	 
 	The eigenvectors have the form 
  	 \begin{equation}
 	\begin{gathered}
 		\{(-1, 0, 0, 0, 0, 0, 0, 0, 0, 0, 0, 0, 0, 0, 0, 1), (0, -1, 0, 0, 0, 0, 0, 0, 0, 0, 0, 0, 0, 0, 1, 0), \\
	(0, 0, -1, 0, 0, 0, 0, 0, 0, 0, 0, 0, 0, 1, 0, 0), (0, 0, 0, -1, 0, 0, 0, 0, 0, 0, 0, 0, 1, 0, 0, 0), \\ (0, 0, 0, 0, -1, 0, 0, 0, 0, 0, 0, 1, 0, 0, 0, 0), (0, 0, 0, 0, 0, -1, 0, 0, 0, 0, 1, 0, 0, 0, 0, 0), \\
	(0, 0, 0, 0, 0, 0, -1, 0, 0, 1, 0, 0, 0, 0, 0, 0), (0, 0, 0, 0, 0, 0, 0, -1, 1, 0, 0, 0, 0, 0, 0, 0), \\
	(1, 0, 0, 0, 0, 0, 0, 0, 0, 0, 0, 0, 0, 0, 0, 1), (0, 1, 0, 0, 0, 0, 0, 0, 0, 0, 0, 0, 0, 0, 1, 0), \\
	(0, 0, 1, 0, 0, 0, 0, 0, 0, 0, 0, 0, 0, 1, 0, 0), (0, 0, 0, 1, 0, 0, 0, 0, 0, 0, 0, 0, 1, 0, 0, 0), \\
	(0, 0, 0, 0, 1, 0, 0, 0, 0, 0, 0, 1, 0, 0, 0, 0), (0, 0, 0, 0, 0, 1, 0, 0, 0, 0, 1, 0, 0, 0, 0, 0), \\    (0, 0, 0, 0, 0, 0, 1, 0, 0, 1, 0, 0, 0, 0, 0, 0), (0, 0, 0, 0, 0, 0, 0, 1, 1, 0, 0, 0, 0, 0, 0, 0).\} 		
 	\end{gathered}  	 
  \end{equation} 	 	
 	
	The eigenvalues of the matrix $	\Gamma$ are 
  	 \begin{equation}
 	\begin{gathered}
 		\{-1, -1, -1, -1, -1, -1, 1, 1, 1, 1, 1, 1, 1, 1, 1, 1 \}			.		
 	\end{gathered}  	 
 	\end{equation} 	 
 
 	The eigenvectors have the form
 	\begin{equation}
 	\begin{gathered}
 	\{(0, 0, 0, 0, 0, 0, 0, 0, 0, 0, 0, 0, 0, -1, 1, 0), (0, 0, 0, 0, 0, 0, 0, -1, 0, 0, 0, 1, 0, 0, 0, 0), \\
	(0, 0, 0, 0, 0, -1, 0, 0, 0, 0, 1, 0, 0, 0, 0, 0), (0, 0, 0, 0, 0, 0, -1, 0, 0, 1, 0, 0, 0, 0, 0, 0), \\
	(0, 0, 0, 0, -1, 0, 0, 0, 1, 0, 0, 0, 0, 0, 0, 0), (0, -1, 1, 0, 0, 0, 0, 0, 0, 0, 0, 0, 0, 0, 0, 0), \\
	(0, 0, 0, 0, 0, 0, 0, 0, 0, 0, 0, 0, 0, 0, 0, 1), (0, 0, 0, 0, 0, 0, 0, 0, 0, 0, 0, 0, 0, 1, 1, 0), \\
	(0, 0, 0, 0, 0, 0, 0, 0, 0, 0, 0, 0, 1, 0, 0, 0), (0, 0, 0, 0, 0, 0, 0, 1, 0, 0, 0, 1, 0, 0, 0, 0), \\
	(0, 0, 0, 0, 0, 1, 0, 0, 0, 0, 1, 0, 0, 0, 0, 0), (0, 0, 0, 0, 0, 0, 1, 0, 0, 1, 0, 0, 0, 0, 0, 0), \\
	(0, 0, 0, 0, 1, 0, 0, 0, 1, 0, 0, 0, 0, 0, 0, 0), (0, 0, 0, 1, 0, 0, 0, 0, 0, 0, 0, 0, 0, 0, 0, 0), \\    (0, 1, 1, 0, 0, 0, 0, 0, 0, 0, 0, 0, 0, 0, 0, 0), (1, 0, 0, 0, 0, 0, 0, 0, 0, 0, 0, 0, 0, 0, 0, 0)\}.		
 	\end{gathered}  	 
 	 \end{equation} 	 

 	Using the \cite{Horn_Johnson} eigenspaces conservation theorem for commuting matrices, and taking into account that all eigenvectors of the matrix $\Phi$ are centrally symmetric or centrally antisymmetric, we write down 4 possible types of eigenvectors of the matrix $\Theta$.
 	 	
 	 \begin{center}
 	 	\begin{equation} \label{theta2_1}
 	 		\begin{gathered}
 	 		\vec{y_1}= (u_{1}, u_{2}, u_{2}, u_{4}, u_{5}, u_{6}, u_{7}, u_{5}, u_{5}, u_{7}, u_{6}, u_{5}, u_{4}, u_{2}, u_{2}, u_{1}),
 	 		\end{gathered}
 	 	\end{equation}
 	 \end{center}
 	 
  	\begin{center}
 	 	\begin{equation} \label{theta2_2}
 	 		\begin{gathered}
 	 		\vec{y_2}= (u_{1}, u_{2}, u_{3}, u_{4}, u_{5}, u_{6}, u_{7}, u_{8}, u_{8}, u_{7}, u_{6}, u_{5}, u_{4}, u_{3}, u_{2}, u_{1}),
 	 		\end{gathered}
 	 	\end{equation}
 	 \end{center}
 	 
 	 \begin{center} 
     	\begin{equation} \label{theta3_1}
		\begin{gathered}
 		\vec{y_3}= (u_{1}, u_{2}, u_{2}, u_{4}, u_{5}, 0, 0, -u_{5}, u_{5}, 0, 0, -u_{5}, -u_{4}, -u_{2}, -u_{2}, -u_{1}),
		\end{gathered}
	\end{equation}
	\end{center}

   	\begin{center}
	\begin{equation} \label{theta3_2}
		\begin{gathered}
		\vec{y_4}= (0, u_{2}, -u_{2}, 0, u_{5}, u_{6}, u_{7}, u_{5}, -u_{5}, -u_{7}, -u_{6}, -u_{5},0, -u_{2}, u_{2}, 0).
		\end{gathered}
	\end{equation}
	\end{center} 
	 
	As a result, all 16 eigenvectors will be represented through these types, namely 6 eigenvectors of the form $\vec{y_1}$, 2 eigenvectors of the form $\vec{y_2}$, 4 eigenvectors of the form $\vec{y_3}$ and 4 eigenvectors of the form $\vec{y_4}$.
	From the central symmetry or central antisymmetry of the eigenvectors of the $\Theta$ transfer matrix, the eigenvalues of the $\Theta$ matrix will coincide with the eigenvalues of two $8 \times 8$ matrices $\Theta_2$ and $\Theta_3$:  
 	 \begin{equation}
 	 	\begin{gathered}
 	 		\Theta_2 (i,j) = \Theta(i,j) + \Theta(i,17-j), \\
 	 		\Theta_3 (i,j) = \Theta(i,j) - \Theta(i,17-j),
 	 	\end{gathered} 
 	 \end{equation}	
 	 where  $i=1,..8, j = 1...8$.
 	 
 	 The matrix $\Theta_2 $ has the form:

 	 \begin{center}
 	 \begin{equation}
 		\Theta_2 = 
 	 	\begin{pmatrix} 
 	 		p q^{2} r^{2} & 0 & 0 &  \frac{p q^{2}}{r^{2}}& p q^{2} & 0 & 0 & p q^{2} \\
 	 		\frac{1}{p} & 0 & 0 &  \frac{1}{p} & \frac{q}{p} & 0 & 0 &  \frac{1}{p r^{2}} \\
 	 		\frac{1}{p} & 0 & 0 & \frac{1}{p} & \frac{1}{p r^{2}} & 0 & 0 & \frac{r^{2}}{p}\\
 	 		\frac{p}{q^{2} r^{2}} & 0 & 0 & \frac{p r^{2}}{q^{2}} & \frac{p}{q^{2}} & 0 & 0 & \frac{p}{q^{2}} \\
 	 		0 & p r^{2}  & \frac{p}{r^{2}}  & 0 & 0 &  p & p & 0 \\
 	 		0 & \frac{q^{2}}{p} & \frac{q^{2}}{p}  & 0 & 0 & \frac{q^{2}r^{2}}{p} & \frac{q^{2}}{p r^{2}} & 0 \\
 	 		0 & \frac{1}{p q^{2}} & \frac{1}{p q^{2}} & 0 & 0 &  \frac{1}{p q^{2} r^{2}} & \frac{r^{2}}{p q^{2}} & 0 \\
 	 		0 & \frac{p}{r^{2}} &  p r^{2}  & 0 & 0 & p & p & 0
 	 	\end{pmatrix}.
  	\end{equation}
 	\end{center}

	Taking into account the form of the eigenvectors (\ref{theta2_1}-\ref{theta2_2}) the characteristic polynomial of the $\Theta_2 $ matrix can be easily factorized:
	\begin{center} 
	\begin{equation} \label{poly_4_1}
		\begin{gathered}
			(-1 + r^4 - r^2 x)^2 \cdot (-1 + r^4 + r^2 x)^2 \cdot 
			(a_{1} x^{4}+b_{1} x^{3} + c_{1} x^{2} + d_{1} x + e_{1}),
		\end{gathered} 
	\end{equation}
	\end{center}
	where 
	$$a_{1}= q^4 r^8,  b_{1} = -p (1 + p^2) q^2 (1 + q^4) r^{10}, $$
	$$c_{1} = -q^4 r^4 - p^4 q^4 r^4 - 4 p^2 q^4 r^8 + p^2 r^{12} + q^4 r^{12} + 2 p^2 q^4 r^{12} + p^4 q^4 r^{12} + p^2 q^8 r^{12}, $$
	$$d_{1} = -p (1 + p^2) q^2 (1 + q^4) (-1 + r)^2 r^6 (1 + r)^2 (1 + r^2)^2,$$
	$$ e_{1} = p^2 q^4 (-1 + r)^4 (1 + r)^4 (1 + r^2)^4.$$
	 
	 The 4th degree polynomial in (\ref{poly_4_1})
	
		\begin{center}
			\begin{equation}
				\begin{gathered}
							a_{1} x^{4}+b_{1} x^{3} + c_{1} x^{2} + d_{1} x + e_{1}
						\end{gathered}
				\end{equation}
			\end{center}
	
	decomposes into a product of square trinomials:	
	\begin{center}
		\begin{equation}
			\begin{gathered}
			(p\cdot q^2\cdot r^4\cdot x^2 + h_{1}\cdot x + 
			p \cdot q^2 (-1 + r)^2 (1 + r)^2 (1 + r^2)^2) \cdot \\
			(p\cdot q^2\cdot r^4 \cdot 
			x^2 + h_{2}\cdot x + p\cdot q^2 (-1 + r)^2 (1 + r)^2 (1 + r^2)^2).
			\end{gathered} 
		\end{equation}
	\end{center}

	In this case, we obtain a system of equations for the parameters $h_{1}$ and $h_{2}$ :
	\begin{equation} \label{system_f}
	\begin{cases}
		x \cdot p q^{2} \cdot (-1 + r)^{2} \cdot (1 + r)^{2} \cdot (1 + r^{2})^{2} \cdot (h_{1} + h_{2} + r^{6} + p^{2} r^{6} + q^{4} r^{6} + p^{2} q^{4} r^{6}) = 0, \\ 
		
		x^{2} \cdot (-h_{2}^{2} + q^{4} r^{4} + 2 p^{2} q^{4} r^{4} + p^{4} q^{4} r^{4} - h_{2} r^{6} - h_{2} p^{2} r^{6} - h_{2} q^{4} r^{6} - h_{2} p^{2} q^{4} r^{6} -  p^{2} r^{12} - \\
		- q^{4} r^{12} - p^{4} q^{4} r^{12} - p^{2} q^{8} r^{12} ) = 0, \\ 
		
		x^{3} \cdot p \cdot q^{2} \cdot r^{4} \cdot (h_{1} + h_{2} + r^{6} + p^{2} r^{6} + q^{4} r^{6} + p^{2} q^{4} r^{6}) = 0.						
	\end{cases}
	\end{equation}

	Note that the first and the third equations of the system (\ref{system_f}) are the same (let us suppose that $r \neq 1$). As a result, we get		
	\begin{equation}
	\begin{cases}
		h_{1} = -\frac{1}{2} (r^6(1+p^2)(1+q^4) - r^2 [4 q^4 + 8 p^2 q^4 + 4 p^4 q^4 + r^8 - \\
		- 2 p^2 r^8 + p^4 r^8 - 2 q^4 r^8 + 4 p^2 q^4 r^8 - 2 p^4 q^4 r^8 + 
		q^{8} r^8 - 2 p^2 q^{8} r^8 + p^4 q^{8} r^8]^{\frac {1} {2}} ), \\ \\
		
		h_{2} =	-\frac{1}{2} (r^6(1+p^2)(1+q^4) + r^2 [4 q^4 + 8 p^2 q^4 + 4 p^4 q^4 + r^8 - \\
		- 2 p^2 r^8 + p^4 r^8 - 2 q^4 r^8 + 4 p^2 q^4 r^8 - 2 p^4 q^4 r^8 + 
		q^{8} r^8 - 2 p^2 q^{8} r^8 + p^4 q^{8} r^8]^{\frac {1} {2}} ) .
	\end{cases}
	\end{equation}

	Let us write the eigenvalues $\lambda_i, i=1,...,8$ of the matrix $\Theta_2$:
	\begin{center}
		\begin{equation}
			\begin{gathered}
				\lambda_{1} =  \frac{-h_{2}+\sqrt{h_{2}^{2}-4 p^{2} q^{4} r^{4}(r^{4}-1)^{2}}} {2 p q^{2} r^{4}},
			\end{gathered}
		\end{equation}
	\end{center}	

	\begin{center}
		\begin{equation}
			\begin{gathered}
				\lambda_{2} = \frac{-h_{2}-\sqrt{h_{2}^{2}-4 p^{2} q^{4} r^{4}(r^{4}-1)^{2}}} {2 p q^{2} r^{4}},
			\end{gathered}
		\end{equation}
	\end{center}	

	\begin{center}
		\begin{equation}
			\begin{gathered}
				\lambda_{3} = \frac{-h_{1}+\sqrt{h_{1}^{2}-4 p^{2} q^{4} r^{4}(r^{4}-1)^{2}}} {2 p q^{2} r^{4}},
			\end{gathered}
		\end{equation}
	\end{center}	

	\begin{center}
		\begin{equation}
			\begin{gathered}
				\lambda_{4} = \frac{-h_{1}-\sqrt{h_{1}^{2}-4 p^{2} q^{4} r^{4}(r^{4}-1)^{2}}} {2 p q^{2} r^{4}}, 
			\end{gathered}
		\end{equation}
	\end{center}	

	\begin{center}
		\begin{equation}
			\begin{gathered}
				\lambda_{5,6} =\frac{r^{4}-1}{r^{2}}, 
			\end{gathered}
		\end{equation}
	\end{center}	

	\begin{center}
		\begin{equation}
			\begin{gathered}
				\lambda_{7,8} =\frac{1-r^{4}}{r^{2}}.  
			\end{gathered}
		\end{equation}
	\end{center}
	
	According to the Perron-Frobenius theorem  for matrices with positive matrix elements, there exists a positive eigenvalue $\lambda_{max}$ such that all other eigenvalues of this matrix are strictly less than its absolute value. This theorem is applicable for a matrix with non-zero positive elements $\Theta^{2}$.  The largest eigenvalue of the matrix $\Theta$ is $\lambda_{1}$.

	\begin{center}
	\begin{equation} 
		\begin{gathered} \label{lambda_max}
			\lambda_{max} = \lambda_{1} = \frac{1}{4p q^{2} r^{4}} (r^6(1+p^2)(1+q^4) + r^2  [(-1+p^2)^2 r^8 + \\(-1 + p^2)^2 q^{8} r^8   
			+ q^{4} (4-2r^8-2p^4(-2+r^8) + 4p^2(2+r^8))]^{\frac {1} {2}} + \\
			+[r^{4} (-16 p^{2} q^{4} + 32 p^2 q^4 r^4 - 16 p^2 q^4 r^8  
			+ ((1 + p^2) (1+q^4) r^4 + \\
			+[(-1 + p^2)^2 r^8 + (-1 + p^2)^2 q^8 r^8 + q^4 (4-2r^8-2p^4(-2+r^8) + 4p^2(2+r^8))]^{\frac {1} {2}}))^2]^{\frac {1} {2}}). 
		\end{gathered}
	\end{equation}
	\end{center}

   The eigenvector corresponding to the largest eigenvalue $\lambda_{1}$ has the form (\ref{theta2_1}).
	The matrix $\Theta_3$ has the form:
	\begin{center}
		\begin{equation} \label{matrix_theta_3}
			\Theta_3 = 
			\begin{pmatrix} 
				p q^{2} r^{2} & 0 & 0 &  -\frac{p q^{2}}{r^{2}}& p q^{2} & 0 & 0 & -p q^{2} \\
				\frac{1}{p} & 0 & 0 &  -\frac{1}{p} & \frac{r^{2}}{p} & 0 & 0 &  -\frac{1}{p r^{2}} \\
				\frac{1}{p} & 0 & 0 & -\frac{1}{p} & \frac{1}{p r^{2}} & 0 & 0 & -\frac{r^{2}}{p}\\
				\frac{p}{q^{2} r^{2}} & 0 & 0 & -\frac{p r^{2}}{q^{1}} & \frac{p}{q^{2}} & 0 & 0 & -\frac{p}{q^{2}} \\
				0 & p r^{2}  & -\frac{p}{r^{2}}  & 0 & 0 &  p & -p & 0 \\
				0 & \frac{q^{2}}{p} & -\frac{q^{2}}{p}  & 0 & 0 & \frac{q^{2} r^{2}}{p} & -\frac{q^{2}}{p r^{2}} & 0 \\
				0 & \frac{1}{p q^{2}} & -\frac{1}{p q^{2}} & 0 & 0 &  \frac{1}{p q^{2} r^{2}} & -\frac{r^{2}}{p q^{2}} & 0 \\
				0 & \frac{p}{r^{2}} &  -p r^{2}  & 0 & 0 & p & -p & 0
			\end{pmatrix}.
		\end{equation}
	\end{center}

	Taking into account the form of the eigenvectors (\ref{theta3_1}, \ref{theta3_2}) of the matrix $\Theta$ the eigenvectors of the $\Theta_3$ have the form:
	
	 	 \begin{center} 
		\begin{equation} \label{theta3_11}
			\begin{gathered}
				\vec{w_1}= (u_{1}, u_{2}, u_{2}, u_{4}, u_{5}, 0, 0, -u_{5}),
			\end{gathered}
		\end{equation}
	\end{center}
	
	\begin{center}
		\begin{equation} \label{theta3_22}
			\begin{gathered}
				\vec{w_2}= (0, u_{2}, -u_{2}, 0, u_{5}, u_{6}, u_{7}, u_{5}).
			\end{gathered}
		\end{equation}
	\end{center} 	 

	This helps to decompose the characteristic polynomial of the $\Theta_3$ matrix into the product of two 4th degree polynomials:	
	
	\begin{center}
	\begin{equation} \label{poly_4_2_3}
		\begin{gathered}
	(a_{2} x^{4}+b_{2} x^{3} + c_{2} x^{2} + d_{2} x + e_{2}) \cdot 
	(a_{3} x^{4}+b_{3} x^{3} + c_{3} x^{2} + d_{3} x + e_{3}) ,
		\end{gathered} 
	\end{equation}
	\end{center}	
	where
	$$a_{2} = q^2 r^8,  b_{2} = p r^{10} - p q^4 r^{10}, 
	c_{2} = q^2 r^4 + p^2 q^2 r^4 - q^2 r^{12} - p^2 q^2 r^{12}, $$
	$$d_{2} = -p r^6 + p q^4 r^6 + 2 p r^{10} - 2 p q^4 r^{10} - p r^{14} + p q^4 r^{14},$$
	$$ e_{2} = p^2 q^2 (-1 + r)^4 (1 + r)^4 (1 + r^2)^4,$$
	$$a_{3} = p^2 q^2 r^8,  b_{3} = p r^{10} - p q^4 r^{10}, 
	c_{3} = q^2 r^4 + p^2 q^2 r^4 - q^2 r^{12} - p^2 q^2 r^{12}, $$
	$$d_{3} = -p r^6 + p q^4 r^6 + 2 p r^{10} - 2 p q^4 r^{10} - p r^{14} + p q^4 r^{14},$$
	$$ e_{3} = q^2 (-1 + r)^4 (1 + r)^4 (1 + r^2)^4.$$
	
	The eigenvalues $\lambda_i, i=9,...,16$ of the matrix $\Theta_3$ are the routes  roots of these two fourth degree polynomials (\ref{poly_4_2_3}). We find them by the Ferrari method described in the appendix \ref{appendix} . 
	
	\section{Conclusion} \label{Conclusion}

	In this article, for the considered Axial Next-Nearest Neighbor Ising model without an external field on a closed chain of spins of width 2 the authors found the exact values of the partition function in a finite strip of length $L$ , free energy, internal energy per node, heat capacity in a finite strip of length $L$ and in the thermodynamic limit at $L \to \infty $. The presence in the Hamiltonian of a model of the interaction of spins located at a distance of 2 from each other determines the specifics of the model. The fact that the action of the transfer matrix is directed along this interaction made it possible, even in a strip of width 2, to capture the features of the phase pattern of the model on an infinite flat lattice. The free energy of the model in the thermodynamic limit is expressed in terms of the root of a quadratic equation of a simple form, which facilitates its analysis.

	\section{Acknowledgments} \label{Gratitude}
	
	The authors are indebted and grateful to professor Walter Selke for helpful remarks. 
	
\section{Appendix} \label{appendix}

\subsection{The Ferrari method}

To find the roots of 4th degree polynomials (\ref{poly_4_2_3}), let us use the Ferrari method (\cite{Cardano}).

Let's write the polynomials (\ref{poly_4_2_3}) in the reduced form 

\begin{center}
	\begin{equation} \label{4_form}
		\begin{gathered}
			x^{4} + g_{1} x^{3} + g_{2} x^{2} + g_{3} x + g_{4} = 0. 
		\end{gathered} 
	\end{equation}
\end{center}	

By the Ferrari method, if $y_{1}$ is a real positive root of the auxiliary equation

\begin{center}
	\begin{equation}
		\begin{gathered}
			y^{3} + A y^{2} + B y + C = 0, 
		\end{gathered} 
	\end{equation}
\end{center}	

where $A=-g_{2}, B= g_{1} g_{3} - 4 g_{4}, C = -g_{1}^{2} g_{4} + 4 g_{2} g_{4} - g_{3}^2$, then the four roots of the original equation (\ref{4_form}) are found as the roots of two quadratic equations:
\begin{center}
	\begin{equation} \label{Ferrari_four}
		\begin{gathered}
			x^{2} + (g_{1}/2)x+ y_{1}/2 = \pm \sqrt{(g_{1}/4 - g_{2} + y_{1})x^{2} + ((g_{1}/2)y_{1} - g_{3})x + (y_{1}^2/4 - g_{4})}, 
		\end{gathered} 
	\end{equation}
\end{center}

where the radical expression on the right side is a perfect square.

Using the Cardano-Vieta formulas (\cite{Tabachnikov}), we write out the real positive root $y_{1}$:
\begin{center} 
	\begin{equation} \label{t_first}
		\begin{gathered}
			y_{1} = \frac{1}{3} (-A + 2\sqrt{A^2 - 3B} \sin[{\frac{1}{3} (\arcsin [{\frac{2A^3-9AB + 27C} {2(A^2 - 3B)^{3/2}}}] + 2 \pi)}]). \\ 
		\end{gathered} 
	\end{equation}
\end{center}

Then the roots of \ref{Ferrari_four} can be found as the roots of two quadratic equation:
\begin{center} 
	\begin{equation} 
		\begin{gathered}
			x^2 + px+ q =0, \\ 
		\end{gathered} 
	\end{equation}
\end{center}
	
where $p=g_{1}/2 + \sqrt{g_{1}^2/4 - g_{2} + y_1}, q=g_{1}/2 + \sqrt{g_{1}^2/4 - g_{4}}$ or $p=g_{1}/2 - \sqrt{g_{1}^2/4 - g_{2} + y_1}, q=g_{1}/2 - \sqrt{g_{1}^2/4 - g_{4}}$.

\end{document}